\documentclass[showpacs]{revtex4}
\usepackage{epsfig}
\def\beq{\begin{equation}}
\def\enq{\end{equation}}
\def\beqa{\begin{eqnarray}}
\def\enqa{\end{eqnarray}}

\def\MeV{\nobreak\,\mbox{MeV}}
\def\GeV{\nobreak\,\mbox{GeV}}

\def\G3{\lag g^3G^3\rag}

\def\nn{\nonumber}

\newcommand{\rag}{\rangle}
\newcommand{\lag}{\langle}

\begin{document}

\title{\sc
Review and interpretation of the new heavy states discovered at the 
$B$-factories
}
\author{Marina Nielsen}
\email{mnielsen@if.usp.br}
\affiliation{Instituto de F\'{\i}sica, Universidade de S\~{a}o Paulo,
C.P. 66318, 05389-970 S\~{a}o Paulo, SP, Brazil}

\begin{abstract}
I discuss some recent discoveries in the spectroscopy of the charmonium-like
resonances, and some of their theoretical interpretations are reviewed.

\end{abstract}

\maketitle

\section{Introduction}

The study of spectroscopy and the decay properties of the heavy flavor 
mesonic states provides us useful information about the dynamics of quarks
and gluons at the hadronic scale. The remarkable progress at the 
experimental side, with various high energy machines, like the $B$-factories,
has opened up new challenges in the theoretical understanding of heavy flavor
hadrons.

The $B$-factories, the PEPII at SLAC in the U.S.A., and the KEKB at KEK in
Japan, were constructed to test the Standard Model mechanism for 
CP violation. They are $e^+e^-$ colliders operating at a CM energy near
10,580 MeV. The $B\bar{B}$ pairs produced are measured by the BaBar (SLAC)
and Belle (KEK) collaborations.

An unexpected bonus of the $B$-factories has been a number of interesting 
contributions to the field of hadron spectroscopy, in particular in the
area of charmonium spectroscopy. At the quark level, the $b$ quark decays
weakly to a $c$ quark accompained by the emission of a virtual $W^-$ boson.
Approximately half of the time, the $W^-$ boson materializes as a $s\bar{c}$
pair. Therefore, half of the $B$ meson decays result in a final state that
contains a $c\bar{c}$ pair. When these $c\bar{c}$ pairs are produced close to
each other in phase space, they can coalesce to form a $c\bar{c}$
charmonium meson.

The simplest charmonium producing $B$ meson decay is: $B\to K(c\bar{c})$.
Another interesting  form to produce charmonium in $B$-factories is
directly form the $e^+e^-$ collision, when the initial state $e^+$ or $e^-$
occasionally radiates a high energy $\gamma$-ray, and the  $e^+e^-$
subsequently annihilate at a corresponding reduced CM energy. When the
energy of the radiated $\gamma$-ray $(\gamma_{ISR})$ is between 4000
and 5000 MeV, the $e^+e^-$ annhilation occurs at CM energies that correspond
to the range of mass of the charmonium mesons. Thus, the initial state
radiation (ISR) process can directly produce charmonium states with
$J^{PC}=1^{--}$.

In the next sections I discuss the experimental data and the possible 
interpretations for the recently observed $X$, $Y$ and $Z$ mesons.

\section{ The $X(3872)$ meson}

In August 2003, Belle reported evidence for a new narrow state in the
decay $B^+\!\rightarrow\!X(3872)K^+\rightarrow\!J/\psi\pi^+\pi^- K^+$
\cite{belle1}, which has been confirmed by CDF, D0 and BaBar \cite{Xexpts}. 
The current world average mass is
\beq
M_X=(3871.4\pm0.6)\MeV\;,
\label{xmass}
\enq
and its total width is less than 2.3 MeV. Belle's \cite{belleE} and BaBar's
\cite{babar2} observation of the decay
$X(3872) \to J/\psi\,\gamma$ determines $C=+$, opposite
to the charge-conjugation of the leading charmonium candidates.  Angular
correlations among the final state particles from $X(3872)\rightarrow
J/\psi\pi^+\pi^-$ decay strongly suggests $J^{PC}=1^{++}$ quantum numbers
\cite{cdf2}.

From constituent quark models \cite{bg} the masses of the possible charmonium
states with $J^{PC}=1^{++}$ quantum numbers are; $2~^3P_1(3990)$ and
$3~^3P_1(4290)$, which are much bigger than the observed mass.

Evidence for the decay  $X(3872) \to J/\psi\,\pi^+\pi^-\pi^0$ at a rate 
comparable to that of $X(3872)\rightarrow\!J/\psi\pi^+\pi^-$ was also
observed by Belle \cite{belleE}: 
\beq
{X \to J/\psi\,\pi^+\pi^-\pi^0\over X\to\!J/\psi\pi^+\pi^-}=1.0\pm0.4\pm0.3.
\label{rate}
\enq
This observation establishes strong isospin and G parity violation, which is
incompatible with a $c\bar{c}$ structure for $X(3872)$.

The observation of these two decays, plus the coincidence between the $X$
mass and the $D^{*0}D^0$ threshold: $M(D^{*0}D^0)=(3871.81\pm0.36)\MeV$
\cite{cleo}, inspired the proposal that the $X(3872)$ could be a molecular
$(D^{*0}\bar{D}^0+\bar{D}^{*0}D^0)$ bound state with small binding energy
\cite{close,swanson}. As a matter of fact, Tornqvist, using a meson potential
 model \cite{torn}, essentially predicted the $X(3872)$ in 1994, since he 
found that there should be molecules near the $D^*D$ threshold in the
$J^{PC}=0^{-+}$ and $1^{++}$ channels. The only other molecular state that is
 predicted in the potential model updated by Swanson is a $0^{++}~~D^*
\bar{D}^*$ molecule at 4013 MeV \cite{swanson}. The $D^{*0}\bar{D}^0$ 
molecule is not an isospin eigenstate and the rate in Eq.(\ref{rate}) is
explained in a very natural way in this model.

Recently Belle \cite{belleD}
and BaBar \cite{babar3} Collaborations reported a near threshold 
enhancement in the $D^0\bar{D}^0\pi^0$ system. The peak mass values for the 
two observations are in good agreement with each other:
$(3875.2\pm1.9)$ MeV for Belle and $(3875.1\pm1.2)$ MeV for BaBar, and are 
higher 
than in the  mass of the $X(3872)$ observed in the $J/\psi\pi^+\pi^-$ 
channel 
by $(3.8\pm1.1)$ MeV. Since this peak lies about 3 MeV above the $D^{*0}
\bar{D}^0$ threshold, it is very ackward to treat it as a $D^{*0}\bar{D}^0$ 
bound state. According to Braaten \cite{braaten}, the peak observed in 
the $B\to K~D^0\bar{D}^0\pi^0$ decay channel is a combination of a resonance 
below the $D^{*0}\bar{D}^0$ threshold from the $B\to K~J/\psi\pi^+\pi^-$ 
decay and a threshold enhancement above the $D^{*0}\bar{D}^0$ threshold.
However, in an updated study \cite{belled*d},
the new value for the mass of the near threshold enhancement 
in the $D^0\bar{D}^0\pi^0$ system reported by the  Belle Collaboration
 is $(3872.6^{+0.5}_{-0.4}\pm0.4)$ MeV, in a
very good agreement with the current world average mass for the $X(3872)$
in the $J/\psi\pi^+\pi^-$ mode in Eq.~(\ref{xmass}).

Maiani {it et al.} \cite{maiani} advocate a tetraquark explanation for the
$X(3872)$. They have considered diquark-antidiquark states with $J^{PC}=
1^{++}$  and symmetric spin distribution:
\beq
X_q=[cq]_{S=1}[\bar{c}\bar{q}]_{S=0}+[cq]_{S=0}[\bar{c}\bar{q}]_{S=1}.
\enq
Physical states could be expected to fall in isospin multiplets with 
$I=0,~1$:
\beq
X(I=0)={X_u+X_d\over\sqrt{2}},\;\;\;\;X(I=1)={X_u-X_d\over\sqrt{2}}.
\enq
However, due to the charm quark mass scale, annihilation diagrams are
suppressed and, therefore, states are closer to mass eigenstates and are no
longer isospin eigenstates. The most general states are:
\beq
X_l=\cos{\theta}X_u+\sin{\theta}X_d,\;\;\;\;X_h=\cos{\theta}X_d-\sin{\theta}
X_u,
\enq
and both can decay into $2\pi$ and $3\pi$. Imposing the rate in 
Eq.(\ref{rate}), they get $\theta\sim20^0$. They also argue that if $X_l$
dominates $B^+$ decays, then $X_h$ dominates the $B^0$ decays and 
vice-versa. Therefore, the $X$ particle in $B^+$ and $B^0$ decays are 
different with  \cite{maiani,polosa}
\beq
M(X_h)-M(X_l)=(8\pm3)\MeV.
\label{2X}
\enq

There are two reports from Belle \cite{belleB0} and Babar  \cite{babarB0}
Collaborations for the observation of the $B^0\to K^0~X$ decay. However, 
these reports are not consistent with each other. While
Belle measures \cite{belleB0}:
\beq
{B^0\to X K^0\over B^+\to X K^+}=0.94\pm0.24\pm0.10,
\enq
and
\beq
M(X)_{B^+}-M(X)_{B^0}=(0.22\pm0.90\pm0.27)\MeV,
\enq
BaBar measures \cite{babarB0}:
\beq
{B^0\to X K^0\over B^+\to X K^+}=0.41\pm0.24\pm0.05,
\enq
and
\beq
M(X)_{B^+}-M(X)_{B^0}=(2.7\pm1.6)\MeV.
\enq

In any case, the mass difference measurements are much larger than the
prediciton in Eq.(\ref{2X}). It is interesting to notice that, using the same
tetraquark structure as in ref.~\cite{maiani}, a QCD sum rule calculation
for the mass difference in Eq.(\ref{2X}) has obtained \cite{x3872}:
\beq
M(X_h)-M(X_l)=(3.3\pm0.7)\MeV,
\enq
in agreement with BaBar measurement. The same calculation \cite{x3872} has 
obtained
\beq
M_X=(3.92\pm0.13)\GeV,
\enq
while a QCD sum rule for the $X(3872)$ resonance considering it as a
$(D^{*0}\bar{D}^0+\bar{D}^{*0}D^0)$ molecular state \cite{lnw} has obtained
\beq
M_X=(3.87\pm0.07)\GeV,
\enq
in a better agreement with the experimental mass. Therefore, from a QCDSR 
point of view, the $X(3872)$ is better described as a $D^*D$ molecular state
than as a diquark-antidiquark state.

To summarize, there is an emerging consensus that the $X(3872)$ is a
multiquark state. In favor of the tetraquark configuration is the existence
of two different states decaying from $B^\pm$ or $B^0$. Therefore, it is very
important the confirmation of the existence of these two states. In favor
of the molecular  configuration is the proximity of the $X(3872)$ mass and
the $D^*D$ threshold.

\section{The $Y(J^{PC}=1^{--})$ family}

The $Y(4260)$ was the first one in the family observed by BaBar 
Collaboration \cite{babar1} in the reaction
\beq
e^+e^-\to\gamma_{ISR}J/\psi\pi^+\pi^-,
\enq
with mass $M=(4259\pm10)\MeV$ and width $\Gamma=(88\pm24)\MeV$. It was
confirmed by CLEO and Belle Collaborations \cite{yexp}. The $\pi\pi$ mass
distribution reported in \cite{babar1} peaks near 1 GeV and this information
was interpreted as consistent with the $f_0(980)$ decay. In a updated
report \cite{babary}, BaBar has confirmed the observation of the $Y(4260)$
with a mass and width
\beq
M_Y=(4252\pm7)\MeV,\;\;\;\; \Gamma_Y=(105\pm20)\MeV.
\enq
However, the new $\pi\pi$ mass distribution shows a more complex structure.

BaBar \cite{babar4} also found a broad peak in the reaction
\beq
e^+e^-\to\gamma_{ISR}\psi^\prime\pi^+\pi^-,
\enq
which was confirmed by Belle \cite{belle4}. Belle found that the $\psi^\prime
\pi^+\pi^-$ enhancement observed by BaBar was, in fact, produced by two
distinct peaks with masses and widths:
\beqa
Y(4360):&&\;\;M=(4361\pm13)\MeV,\;\;\;\; \Gamma=(74\pm18)\MeV,
\nn\\
Y(4660):&&\;\;M=(4664\pm12)\MeV,\;\;\;\; \Gamma=(48\pm15)\MeV.
\enqa

The masses and widths of these three states are not consistent with any of
the established $1^{--}$ charmonium states \cite{zhu}, and they can also be
candidates for multiquark states or charmonium hybrids \cite{gool}. An 
attractive interpretation is that the $Y(4260)$ is a charmonium hybrid. 
Hybrids are hadrons in which the gluonic degree of freedom has been excited.
The nature of this gluonic excitation is not well understood, and has been
described by various models. The
spectrum of charmonium hybrids has been calculated using lattice gauge theory
\cite{lattice}. Their result for the mass is approximately 4200 MeV, which
is consistent with flux tube model predictions \cite{bsc}. However, more 
recent lattice simulations predict that the lightest
charmonium hybrid is about 4400 MeV \cite{latticenew} , which is closer
to the mass of the $Y(4360)$.

A critical information for understanding the structure of these states is 
wether
the pion pair comes from a resonance state. From the di-pion invariant mass 
spectra shown in ref.~\cite{babarconf} there is some indication that only the
$Y(4660)$ has a well defined intermediate state consistent with $f_0(980)$
\cite{babarconf}. Due to this fact and the proximity of the mass of the
$\psi'-f_0(980)$ system with the mass of the $Y(4660)$ state, in 
ref.~\cite{ghm},
the $Y(4660)$ was considered as a $f_0(980)~\psi'$ bound state. The $Y(4660)$
was also suggested to be a baryonium state \cite{qiao} and a canonical  
5 $^3$S$_1$ $c\bar{c}$ state \cite{dzy}.

In the case of $Y(4260)$, in ref.~\cite{maiani2} it was considered as a 
$sc$-scalar-diquark $\bar{s}\bar{c}$-scalar-antidiquark in a $P$-wave state. 
Maiani {\it et al.} \cite{maiani2} tried different ways to determine the 
orbital term and they arrived at $M=(4330\pm70)\MeV$, which is more consistent
with $Y(4360)$. However, from the $\pi\pi$ mass 
distribution in ref.~\cite{babarconf}, none of these two states,  $Y(4260)$
and $Y(4360)$ has a decay
with a intermediate state consistent with $f_0(980)$ and, therefore, it is 
not clear that they should have an $s\bar{s}$ pair in their structure.
Also, in ref.~\cite{efg}, using a relativistic diquark-antidiquark picture,
it was shown that the $Y(4260)$ can not be interpreted as a ($[sc]_{S=0}
[\bar{s}\bar{c}]_{S=0}$) state in a $P$-wave. 

If one looks at the threshold of the mesonic systems: $M(D(1865)\bar{D}_1
(2420))\sim 4285\MeV$
and $M(D_0(2310)\bar{D}^*(2007))\sim 4320\MeV$, which have $J^{PC}=1^{--}$ in
$S$-wave, one sees that a molecular interpretation is also possible for 
$Y(4260)$ and $Y(4360)$. In refs.~\cite{rapha,z12} a QCD sum rule calculation
for these molecular states was considered. The obtained mass for the
$D_0\bar{D}^*$ state was: $m_{D_0\bar{D}^*}=(4.27\pm0.10)\GeV$ in good 
agreement with the $Y(4260)$ mass. In the case of the $D\bar{D}_1$ molecular
state, the obtained mass was: $m_{D\bar{D}_1}=(4.19\pm0.22)\GeV$. Therefore,
considering the errors and the width of the $Y(4260)$ meson, the
molecular $D\bar{D}_1$ assignement is also possible, in agreement with the
findings of ref.~\cite{ding}, where a meson exchange model was used to
study the $Y(4260)$ meson.

The authors of ref.~\cite{rapha} also considered diquark-antidiquark states 
with $J^{PC}=1^{--}$  and symmetric spin distribution:
\beq
Y_q=[cq]_{S=1}[\bar{c}\bar{q}]_{S=0}+[cq]_{S=0}[\bar{c}\bar{q}]_{S=1},
\enq
with $q$ standing for a light or a strange quark. The obtained masses were:
$m_{Y_u}=(4.49\pm0.11)\GeV$ and $m_{Y_s}=(4.65\pm0.10)\GeV$. Therefore, the
authors concluded that it is possible to interpret the $Y(4660)$ meson as
a $[cs][\bar{c}\bar{s}]$ diquark-antidiquark state, and this is consistent 
with the di-pion invariant mass spectra shown in ref.~\cite{babarconf} for
$Y(4660)$, since there is some indication that it has a well defined 
intermediate state consistent with $f_0(980)$.

To summarize, the discovery of the $Y(4260),~Y(4360)$ and $Y(4660)$ appears
to represent an overpopulation of the expected charmonium $1^{--}$ states.
The absence of open charm production is also inconsistent with a conventional
$c\bar{c}$ explanation. Possible explanations for these states include 
charmonium hybrid and $D_0\bar{D}^*$ or $D\bar{D}_1$ molecular state for 
$Y(4260)$, charmonium hybrid  and $[cs]_{S=0}[\bar{c}\bar{s}]_{S=0}$
in a $P$-wave tetraquark state for $Y(4360)$, and
a symmetrical $[cs]_{S=1}[\bar{c}\bar{s}]_{S=0}$ tetraquark state or a 
canonical  5 $^3$S$_1$ $c\bar{c}$ state for $Y(4660)$. 
The current situation regarding the $1^{--}$ states produced via ISR is 
clearly unsettled.

\section{The $Z^+(4430)$ meson}

All states discussed so far are electrically neutral. The real turning point
in the discussion about the structure of the new observed charmonium states
was the observation by Belle Collaboration of a charged state decaying into 
$\psi'\pi^+$, produced in $B^+\to K\psi'\pi^+$ \cite{bellez}. 
The measured mass
and width of this state is $M=(4433\pm5)\MeV$, $\Gamma=(45^{+39}_{-18})\MeV$.
There are no reports of a $Z^+$ signal in the $J/\psi\pi^+$ decay channel.
Since the minimal quark content of this state is $c\bar{c}u\bar{d}$, this
state is a prime candidate for a multiquark meson. Since $Z^+(4430)$ was
observed in the $\psi'\pi^+$ channel, it is an isovector state with positive 
$G$-parity: $I^G=1^+$.

There are many theoretical interpretations for the $Z^+(4430)$ structure.
Because its mass is close to the $D^*D_1$ threshold, Rosner \cite{rosner}
suggested it is an $S$-wave threshold effect, while others considered it
to be a strong candidate for a $D^*D_1$ molecular state \cite{meng,nos,zhuz}.
Other possible interpretations are tetraquark state \cite{maiani3}, or a cusp 
in the $D^*D_1$ channel \cite{bugg}. The tetraquark hypothesis implies that
the $Z^+(4430)$ will have neutral partners decaying into $\psi'\pi^0/\eta$.

Considering the  $Z^+(4430)$ as a loosely bound $S$-wave $D^*D_1$ molecular 
state, the allowed angular momentum and parity are $J^P=0^-,~1^-,~2^-$, 
although the $2^-$  assignment is probably suppressed in the $B^+\to Z^+K$ 
decay by the small phase space. Among the remaining possible $0^-$ and $1^-$ 
states, the former will be more stable as the later can also decay to
$DD_1$ in $S$-wave. Moreover, one expects a bigger mass for the $J^P=1^-$
state as compared to a $J^P=0^-$ state.

In ref.~\cite{nos} the QCD sum rules were used to study the $Z^+(4430)$
considered as a $D^*D_1$ molecular state with $I^G~J^P=1^+~0^-$. The mass
obtained was $M_{Z^+}=(4.40\pm0.10)\GeV$ in an excelent agreement with the
experimental mass. To check if the $Z^+(4430)$ could also be described as
a diquark-antidiquark state, in ref.~\cite{z10} different currents were
considered with $J^P=0^-$ and $~1^-$. The results obtained were:
$M_{Z}(0^-)=(4.52\pm0.09)\GeV$ and $M_{Z}(1^-)=(4.84\pm0.14)\GeV$. From these
results we conclude that while it is also possible to describe the $Z^+(4430)$
as a  diquark-antidiquark state with $J^P=0^-$, the $J^P=1^-$ configuration
is disfavored.

Summarizing, the only open options for the $Z^+(4430)$ structure are 
tetraquark, molecule and  threshold effect. It is important to mention
that during this conference it
was related \cite{belleconf} that BaBar claim no significant evidence for the
existence of the $Z^-(4430)$ in the decay $B^{-,0}\to \psi'\pi^-K^{0,+}$
\cite{z-}. Therefore, a confirmation of the
existence of the $Z^\pm(4430)$ is critical before a complete picture
can be drawn. 

\section{The $Z_1^+(4050)$ and $Z_2^+(4250)$ mesons} 
The $Z^+(4430)$ observation motivated studies of other $\bar{B}^0\to K^-\pi^+
(c\bar{c})$ decays. In particular, the Belle Collaboration has recently
reported the observation of two resonance-like structures in the $\pi^+
\chi_{c1}$ mass distribution \cite{belle3}. The significance of each of the
$\pi^+\chi_{c1}$ structures exceeds 5$\sigma$ and, if they are interpreted
as meson states, their minimal quark content must be $c\bar{c}u\bar{d}$.
They were called $Z_1^+(4050)$ and $Z_2^+(4250)$, and their masses and widths
are $M_1=(4051\pm14^{+20}_{-41})~\MeV$, $\Gamma_1=82^{+21+47}_{-17-22}~\MeV$,
$M_2=(4248^{+44+180}_{-29-~35})~\MeV$, $\Gamma_2=177^{+54+316}_{-39-~61}~
\MeV$. Since they were observed in the $\pi^+\chi_{c1}$ channel,
the only quantum numbers that are known about them are $I^G=1^-$.

Due to the
closeness of the $Z_1^+(4050)$ and $Z_2^+(4250)$ masses to the
$D^{*}\bar{D}^*(4020)$ and $D_1\bar{D}(4285)$ thresholds, these states could 
also be interpreted as molecular states or threshold effects. Lie et al. 
\cite{llz}, using a meson exchange model find strong attraction for the
$D^{*}\bar{D}^*$ system with $J^{P}=0^+$. They conclude that, if future 
experiments confirm the $Z_1^+(4050)$ existence, then it is probably a
$D^{*}\bar{D}^*$ loosely bound molecular state. However, it is very difficult
to understand a bound molecular state which mass is above the $D^{*}\bar{D}^*$
threshold.

In a recent work \cite{z12}, the QCD sum rules formalism was used to study
the $D^{*}\bar{D}^*$ and $D_1\bar{D}$ molecular states with $I^GJ^P=1^-0^+$
and $1^-1^-$ respectively. The mass obtained for these molecular states are:
$M_{D^*D^*} = (4.15\pm0.12)~\GeV$, and $M_{D_1D} = (4.19\pm0.22)~\GeV$.
In ref.~\cite{width}
it was found that the inclusion of the width, in the phenomenological side
of the sum rule, increases the obtained mass for molecular states. This means 
that the introduction of the width in the sum rule calculation, increases 
the mass of the  $D^*\bar{D}^*$ and $D_1\bar{D}$ molecules.  As a result, 
the mass of 
the $D_1\bar{D}$ molecule will be closer to the
observed $Z^+(4250)$ mass, and the mass of the $D^*\bar{D}^*$ molecule will 
be far  from the $Z^+(4050)$ mass. Therefore, the authors of ref.~\cite{z12} 
conclude that it
is possible to describe the $Z_2^+(4250)$ resonance  structure as a
$D_1\bar{D}$ molecular state with $I^GJ^P=1^-1^-$ quantum numbers, and that
the $D^{*}\bar{D}^{*}$ state is probably a virtual state that is not
related with the $Z_1^+(4050)$ resonance-like structure. Considering the 
fact that the $D^*D^*$ threshold (4020) is so close to the $Z_1^+(4050)$ 
mass and that
the $\eta^{\prime\prime}_c(3^1S_0)$ mass is predicted to be around 4050 MeV
\cite{gool}, it is probable that the $Z_1^+(4050)$ is only a threshold
effect \cite{gool}.   

\section{Other multiquark states}

If the mesons $X(3872)$, $Z^+(4430)$, $Y(4260)$ and $Z_2^+(4250)$ are really
molecular states, then many other molecules should exist. A systematic study
of these molecular states and their experimental observation would confirm
its structure and provide a new testing ground for QCD within multiquark
configurations. In this context, a natural extension would be to probe the
strangeness sector. In particular, in analogy with the  meson $X(3872)$,
a $D_sD^*$ molecule with $J^{P}=1^{+}$ could be formed in the $B$ meson decay
$B\to\pi X_s\to\pi(J/\psi K\pi)$. Since it would decay into $J/\psi K^*\to J/
\psi K\pi$, it could be easily reconstructed.

In ref.~\cite{lnw} the QCD sum rules approach was used to predict the mass of 
the $D_sD^*$  molecular state. Such prediction 
is of particular importance for new upcoming experiments
which can investigate with much higher precision the charmonium energy regime,
like the PANDA experiment at the antiproton-proton facility at FAIR, or a
possible Super-B factory experiment. Especially PANDA can do a careful scan of
the various thresholds being present, in addition to precisely going through
the exact form of the resonance curve. The obtained mass was:
$M_{D_sD^*}=(3.97\pm0.08)\GeV$ very close to the $D^*D(3980)$ threshold, and
about 100 MeV bigger than the $X(3872)$ mass. This finding strongly suggests
the possibility of the existence of a $D_sD^*$ molecular state with $J^P=1^+$.

Finally, considering that it was already observed the double-charmonium
 production in the reaction \cite{bellecc}
\beq
e^+e^-\to J/\psi+X(3940),
\enq
it seems that it would be possible the formation of the tetraquark $[cc]
[\bar{u}\bar{d}]$.  Such state with quantum numbers $I=0, ~J=1$ and $P=+1$ 
which, following ref.\cite{ros},
we call $T_{cc}$, is especially interesting. As already noted previously
\cite{ros,zsgr}, the $T_{cc}$ state cannot decay strongly
or electromagnetically into two $D$ mesons in the
$S$ wave due to angular momentum conservation nor in $P$ wave due to
parity conservation. If its mass is below the $DD^*$ threshold, this decay
is also forbidden, and this state would be very narrow.

The most attractive light antidiquark is expected to be the in the
color triplet, flavor anti-symmetric and spin 0 channel. Therefore, 
a constituent quark picture for $T_{cc}$
would be a light anti-diquark in color triplet, flavor  anti-symmetric and
spin 0 ($\epsilon_{abc}[\bar{u}_b\gamma_5 C\bar{d}_c^T]$) combined
with a heavy diquark of spin 1 ($\epsilon_{aef}[c_e^TC\gamma_\mu
c_f]$).  Although the spin 1 configuration is repulsive, its strength is much
smaller than that for the light diquark due to the heavy charm quark mass.
This is why one does not expect a bound $T_{ss}$.

A QCD sum rule for such state gives \cite{tcc}: $M_{T_{cc}}=(4.0\pm0.2)\GeV$
in a very good agreement with the predictions based on the one gluon exchange 
potential model \cite{ros}, and color-magnetic model \cite{zhutcc}.

\section{Final Comments}

As a final remark, it is very important to find experimentally observable
quantities which are sensitive to the quark content of the resonances. In 
ref.~\cite{mprs}, Maiani et al. have shown that the nuclear modification
factor, $R_{CP}$, defined as the ratio between the cross sections in central 
and peripheral collisions between relativistic heavy ions,
can be used for this goal. They have shown that there is a large difference
between the $R_{CP}$ for the $f_0(980)$ produced, in $Au~+~Au$ collisions at 
RHIC, when the $f_0(980)$ is assumed to be a 
four-quark state or a quark-antiquark meson. However, it will be very
difficult to observe the new charmonium states discussed above
in relativistic heavy ions collisions at LHC.
Therefore, it is very important to find out other experimentally observable
quantities which could be sensitive to the quark content of the resonances.

\section*{Acknowledgements}
{This work has been partly supported by FAPESP and CNPq-Brazil.}


\end{document}